# MULTI-SCALE UNCERTAINTY QUANTIFICATION IN GEOSTATISTICAL SEISMIC INVERSION

## SEISMIC INVERSION

Leonardo Azevedo[1]*, Vasily Demyanov[2]

[1]CERENA, DECivil, Instituto Superior Técnico, Lisbon University, Av. Rovisco Pais,

1049-001 Lisbon, Portugal: leonardo.azevedo@tecnico.ulisboa.pt

[2]Institute of Petroleum Engineering, Heriot-Watt University, Riccarton, EH144AS, UK:

v.demyanov@hw.ac.uk

*Corresponding Author:

**ABSTRACT**

Geostatistical seismic inversion is commonly used to infer the spatial distribution of the subsurface petro-elastic properties by perturbing the model parameter space through iterative stochastic sequential simulations/co-simulations. The spatial uncertainty of the inferred petro-elastic properties is represented with the updated a posteriori variance from an ensemble of the simulated realizations. Within this setting, the large-scale geological (metaparameters) used to generate the petro-elastic realizations, such as the spatial correlation model and the global a priori distribution of the properties of interest, are assumed to be known and stationary for the entire inversion domain. This assumption leads to underestimation of the uncertainty associated with the inverted models. We propose a practical framework to quantify uncertainty of the large-scale geological parameters in seismic inversion. The framework couples geostatistical seismic inversion with a stochastic adaptive sampling and Bayesian inference of the metaparameters to provide a more accurate and realistic prediction of uncertainty not restricted by heavy assumptions on large-scale geological parameters. The proposed framework is illustrated with both synthetic and real case studies. The results show the ability retrieve more reliable acoustic impedance models with a more adequate uncertainty spread when compared with conventional geostatistical seismic inversion techniques. The proposed approach separately account for geological uncertainty at large-scale (metaparameters) and local scale (trace-by-trace inversion).

# INTRODUCTION

Reliable uncertainty quantification is vital for making predictions about the subsurface petro-elastic properties as they guide the subsequent reservoir development decisions. Early stages of reservoir lifetime, when there are only a few wells drilled, are associated with the largest uncertainty about the reservoir's properties that needs to be accounted for in modelling. The description of the spatial distribution of the reservoir's internal properties at this stage is mostly based on three-dimensional seismic reflection and limited well-log data. The lack of exhaustive direct measurements of the properties of interest, along with the indirect relationship between seismic reflection data and the subsurface geology, still leaves a large range of uncertainty for large-scale geological model properties, such as facies proportions, spatial correlation and anisotropy for the subsurface property of interest.

The way uncertainty is accounted when seismic reflection and well-log data are integrated within reservoir modeling becomes a key step that would impact hydrocarbon reservoir characterization projects. Traditional workflows that offer various ways of integrating seismic reflection and well-log data into statistical reservoir model through conditioning often limit uncertainty to a given aspect (e.g. intrinsic local heterogeneity) and may be anchored on a unique geological interpretation and/or modelling technique.

Accurate uncertainty quantification requires to account coherently for: (i) uncertainty in seismic data themselves due to the measurement errors, noise and the band-limited nature of seismic reflection data (Tarantola 2005); (ii) the way the seismic data are integrated into the geological model – the elastic model; and (iii) the uncertainty of the geological model itself associated with the large-scale geological parameters (or metaparameters), as for example variograms used to describe the spatial continuity pattern of the elastic property of interest (Malinverno and Briggs (2005); Thore (2015); Phelps et al. 2018).

A common technique to combine seismic reflection and well data within the geo-modelling workflow is, for example, to condition the porosity model to experimental data from the wells and secondary data derived from the seismic reflection data with geostatistical simulations (Doyen 2007). Secondary conditioning data is often based on acoustic and/or elastic impedances derived from the seismic reflection data using seismic inversion techniques. Seismic inversion enables to make inferences about the subsurface geology, in terms of its elastic properties, acoustic and/or elastic impedance or P- and S-wave velocities, given the recorded seismic data acquired from that particular area (Tarantola 2005).

Seismic inversion is a nonlinear and ill-conditioned problem with non-unique solution due the nature of the seismic method itself: the limited bandwidth and resolution of the seismic data, noise, measurement errors and physical assumptions about the involved physical system (Tarantola 2005). Consequently, there are many subsurface elastic models that can return similar seismic responses close to the observed one. Consequently, there is always a variable degree of uncertainty intrinsic to any inverted elastic model.

Also, high correlation coefficients achieved between the synthetic and the observed seismic data does not ensure that the inverted elastic model is close to the real geology due to the non-unique solution of the seismic inversion problem. In fact, the inversion procedure may be trapped in a local minimum far from the global solution resulting in elastic models considerable different from the real subsurface geology.

This paper addresses the issue of metaparameter uncertainty in seismic inversion as illustrated in Thore (2015).Thore (2015) rightly stated that a set of stochastic realizations retrieved from stochastic inversion tend to underestimate the uncertainty in seismic inversion that can be related to the metamodel parameters: wavelet, correlation mode, signal–to–noise level, regularization coefficient. Here we deal we variogram correlation range and the global a prior soft conditioning property distribution as well. The latter is largely uncertainty due to data sparsity and can be biased to the preferential well placement. Malinverno and Briggs (2005) considered two Bayesian approaches to infer the metaparameter uncertainty, namely the naïve Bayes and the hierarchical Bayes. Both gave identical results in their particular case of 1D seismic trace inversion and for the considered uncertainty parameters, however this may not be the case in other applications. In our work, we suggest a more pragmatic approach although rigorously Bayesian that ties well with the conventional geostatistical seismic inversion workflows based on multiple stochastic realizations.

We demonstrate how to combine uncertainty associated with the seismic inversion at a local scale with uncertainty related to larger scale geological features, namely the proportion and the spatial continuity pattern of each facies as interpreted from the existing direct measurements, i.e. well-log data. This work proposes a multiscale uncertainty quantification approach based on iterative geostatistical seismic inversion and Bayesian inference of geological parameters with multiple reservoir model realizations. In light of the mentioned earlier Malinverno and Briggs (2005) formalism, we use a hierarchical approach that combines the Bayesian inference of the metaparameters to account for large-scale geological uncertainty and the traditional geostatistical seismic inversion to account for local small-scale geological uncertainty. The latter can be referred to as an inherent uncertainty as is represented by seeded stochastic realizations.

We will demonstrate the application of the proposed method on a synthetic case, where the true reservoir property distribution is available, and on a real reservoir case where we conduct blind validation tests using existing wells not considered to directly constrain the seismic inversion procedure.

## BACKGROUND

**Seismic Inversion**

Seismic inversion aims to infer the spatial distribution of the petro-elastic properties of the subsurface from the recorded seismic reflection data. Seismic inversion problem can be posed in two different frameworks: a deterministic (or optimization), or a statistical one (Bosch et al. 2010).

Deterministic approaches rely on optimization techniques and result in a single best-fit inverse model, corresponding to the maximum a posteriori, and are always a smooth representation of the real subsurface geology (Bosch et al. 2010). Statistical inversion methods can be distinguished between those that use a Bayesian approach or are based on the stochastic perturbation of the model parameter space (i.e., using stochastic sequential simulation (Deustch and Journel 1998)).

In Bayesian inversion approaches, when the forward model is not linearized the posterior distribution can be estimated using Markov Chain Monte Carlo (MCMC) methods as for example the Mestropolis-Hasting algorithm (Bosch et al. 2009; Gunnin and Glinksy 2004; Anandaroop et al. 2013; Ely et al. 2018). On the other hand, Bayesian linearized inversion (Buland and Omre 2003, Gallop 2006; Grana and Della Rossa 2010; Dubreuil-Boisclair et al. 2012; Amaliksen 2014; Grana 2016; Grana et al. 2017; Grana 2018; Fjeldstad and Grana 2018; de Figueiredo et al. 2018; Lang and Grana 2018) takes advantage of the linearization of the forward-model operator and by assuming Gaussian, or a Gaussian mixture, for both the prior probability distributions of the petro-elastic properties of interest and the error associated with the recorded seismic reflection data.

Within these assumptions, the posterior probability distribution for each petro-elastic properties of interest, is analytically expressed in terms of Gaussian, or a Gaussian mixture, distributions.

The resulting posterior probability distribution functions, computed individually for each petro-elastic property inferred from the seismic data, represent the spatial uncertainty related to estimated value. In this context, the uncertainty quantification is limited to the spatial distribution of these properties and does not include any assumption regarding the geological parameters (e.g., variogram models describing the spatial distribution of the inverted property).

Stochastic approaches explore the model parameter space based on the Monte Carlo rejection sampling technique avoiding the Gaussian assumptions of Bayesian linearized seismic techniques. These benefits are achieved with an increase on the computational cost of the inversion procedure (Bosch et al. 2010; Azevedo and Soares 2017).

Geostatistical seismic inversion algorithms, a particular case within stochastic seismic inversion techniques, have increased considerably their importance in reservoir modeling and characterization due to their effectiveness in integrating simultaneously the seismic reflection and the well-log data to retrieve high-resolution subsurface models (Doyen 2007; Azevedo and Soares 2017). Geostatistical seismic inversion methodologies, as introduced by Bortolli et al. (1992) and Hass and Dubrulle (1994), are trace-by-trace inversion methodologies based on genetic algorithms, where the model parameters space is perturbed recurring to stochastic sequential simulation and co-simulation (Deutsch and Journel 1998). Later, Soares et al. (2007) introduced global geostatistical seismic inversion methodologies, where the model parameter space (i.e., the inversion grid) is perturbed at once at each iteration generating an ensemble of elastic models at each iteration (Soares et al. 2007; Nunes et al. 2012; Azevedo et al. 2018). When comparing against trace-by-trace techniques, global iterative geostatistical seismic inversion methodologies are able to avoid fitting the inverted seismic reflection to low signal-to-noise areas within the recorded seismic reflection data (Soares et al. 2007). These areas will remain unmatched throughout the inversion procedure and will be associated with higher variability within the resulting ensemble of inverted elastic models.

The overall framework for global iterative geostatistical seismic inversion methodologies can be described by the following sequence of steps:

i) A set of impedance models is created for the entire inversion grid using a stochastic sequential simulation technique (e.g. Sequential Gaussian Simulation (Deutsch and Journel 1998) or Direct Sequential Simulation (DSS; Soares, 2001));

ii) A synthetic seismic volume is calculated for each of the simulated impedance models and compared on a trace-by-trace basis against the corresponding real trace;

iii) The elastic traces from the ensemble of impedance models generated in i), that produce the synthetic traces that best correlate with the real seismic form an auxiliary grid volume along with the trace-by-trace correlation coefficients;

iv) At the next iteration, a new set of impedance models is generated by stochastic sequential co-simulation using the auxiliary volumes with the best elastic traces and corresponding correlation coefficients, created at the end of the previous iteration, as secondary variables;

v) Return to ii) and iterate until a given global correlation coefficient reaches a certain threshold.

The uncertainty represented in the inverted elastic models resulting from iterative geostatistical inversion approaches is intrinsically associated with the random path followed by the sequential simulation algorithm embedded as part of the inversion technique. The elastic value generated at a certain grid node location depends on the random path, which defines the sequence of conditioning data, composed by the available experimental data and previously simulated nodes are introduced to the model. We can define this uncertainty as being local and inherent to that particular location at the simulation grid.

In addition, seismic inversion techniques that use stochastic sequential simulation as the model perturbation technique ensure the reproduction of the spatial continuity pattern. This is revealed by a variogram model, applied within the given neighborhood, and the prior probability distributions for the properties of interest as estimated from the existing well-log data. These properties come with the assumption of stationarity for both the probability distributions and the imposed spatial continuity pattern: both remain constant during the entire inversion procedure and are considered valid for the entire inversion area. Consequently, the retrieved inverse models do not account for uncertainty in these parameters that are related to large-scale geological heterogeneity, as for example, lateral changes on the sedimentary environment of the study area not captured by the existing experimental data.

This is hardly realistic in real case studies, and especially, for fields at early life stages with only a few drilled wells. In such cases, modeling the horizontal variogram ranges is hardly reliable due to sparsity of the available well data. Moreover, the prior probability distribution as estimated from the well-log is often biased since most wells are drilled in sand-prone areas and, therefore, are not representative of the less porous lithologies. This bias is usually removed by updating the prior probability distribution with the analogue data and geological expert knowledge, which introduces additional uncertainty. Therefore, the uncertainty about the spatial correlation (variogram model) and especially the facies proportions (away from the wells) still remains and needs to be inferred.

Assessing different levels of uncertainty and its joint interpretation with the retrieved subsurface inverse models allows for better risk assessment leading to better decision making. Moving from simpler to more complex and highly heterogeneous hydrocarbon reservoirs the uncertainty assessment of the imposed geological model is of outmost importance and widely used simple perturbations of the property model using only sequential stochastic simulated realizations are not enough.

**Uncertainty Quantification**

Uncertainty quantification with adaptive stochastic sampling algorithms is often used to explore the model parameter space, by generating multiple geological scenarios, to find high likelihood models. Iterative stochastic algorithms (e.g. Neighborhood Algorithm (NA; Sambridge 1999), particle swarm optimization, differential evolution (Hajizadeh et al. 2011) are commonly used for this purpose. The model parameters (e.g., large-scale geological parameters) get updated at each iteration driven by a misfit (*M*) calculated between observed and simulated data.

Uncertainty quantification of global large-scale geological metaparameters, explored by these stochastic sampling techniques, can be inferred in a Bayesian setting. The posterior probability distribution (PPD) functions for each parameter, represent the uncertainty of the geological parameters considered. Coupling this framework with stochastic seismic inversion techniques allows to complement the spatial uncertainty of the inverted elastic properties as inferred by conventional seismic inversion approaches. The PPD and can be approximated given the observed seismic data following Bayes theorem:

$$p(\boldsymbol{m}|\boldsymbol{o}) = \frac{p(\boldsymbol{o}|\boldsymbol{m})p(\boldsymbol{m})}{p(\boldsymbol{o})},\tag{1}$$

where $p(\boldsymbol{m}|\boldsymbol{o})$ is the posterior probability for each parameter considered, given the seismic reflection data $\boldsymbol{o}$; $p(\boldsymbol{o}|\boldsymbol{m})$ is the likelihood function, which is the probability of the seismic reflection data $\boldsymbol{o}$, given the model $\boldsymbol{m}$ is true; $p(\boldsymbol{m})$ is the prior probability and $p(\boldsymbol{o})$ is the evidence computed as the normalization constant.

A general Bayes rule equation (1) has a normalized form, assuming a linear data space and negligible observation uncertainties (Tarantola 2005):

$$p(\boldsymbol{m}|\boldsymbol{o}) = \frac{p(\boldsymbol{o}|\boldsymbol{m})p(\boldsymbol{m})}{\int p(\boldsymbol{o}|\boldsymbol{m})p(\boldsymbol{m})dm}.\tag{2}$$

The uncertain model $\boldsymbol{m}$ in equation (2) is covered by the set of metaparameters. There are several ways of resolving the integral in equation (2) to compute the posterior inference. An analytical solution is possible for equation (2) given the likelihood and prior are defined analytically (e.g., a Gaussian distribution, which is rarely the case). Malinverno and Briggs (2005) recall two approaches – the naïve Bayes and the hierarchical Bayes. The naïve Bayes fixes the metaparameter value at the maximum of the posterior (MAP), which is computationally simple but may tend to underestimate the uncertainty especially in case of a large number of metaparameters. The hierarchical Bayes approximates the full PPD with MCMC, which is accurate enough with a large number of samples. The latter often becomes a computational burden.

In this work we inferred the PPD for each metaparameter through adaptive stochastic sampling in high-dimensional metaparameter space and computed an approximation using Gibbs sampling for the posterior. The approach is computationally feasible since adaptive stochastic sampling concentrates only on the high likelihood regions of the parameter space, which reduces the number of samples. The Gibbs sampling computes the normalizing integral over a high dimensional PDD – the denominator in equation (2). The PPD approximation was computed using NA-Bayes algorithm (Sambridge 1999) which implement a Gibbs sampling on a proxy likelihood surface represented by Voronoi cells, centered around the sampled models in the parameter space. The likelihood within each Voronoi cell is assumed constant and equal to the one computed at the sampling stage for the corresponding model in the center of the cell. NA-Bayes performs resampling with the probability of accepting the Metropolis step, which is proportional to the volume of the Voronoi cell and likelihood.

Within the proposed framework, the prior probability distribution for each of the uncertain parameters is defined based on our beliefs about the subsurface geology of the study area and may be inferred from analog fields or expert knowledge. These prior probability distributions reflect the uncertainty related to each parameter individually.

The likelihood $p(o|m)$ in equation (1) is estimated based on the match of the trace-by-trace synthetic seismic reflection, resulting at each step of the iterative inversion procedure, to the real seismic. A standard likelihood model assumes Gaussian errors, i.e. the errors are independent and the misfit defined by the least square misfit score ($M$):

$$p(\boldsymbol{m}|\boldsymbol{o}) \sim e^{-M}. \qquad\qquad (3)$$

In this work we used a modified misfit score $M$ to determine the closeness of the trace-by-trace match based on the correlation coefficient between the real and synthetic seismic volumes ($CC_{real-synthetic}$) (Equation 4):

$$M = \sum_{i=1}^{n} \frac{1 - CC_{real-synthetic}}{2\sigma^2}, \qquad\qquad (4)$$

where $n$ is the number of nodes (or CMP locations) in the inversion grid and $\sigma^2$ represents the interval of confidence within which we consider a good match in terms of trace-by-trace correlation coefficient between real and synthetic seismic reflection data. This is not exactly the lease squares norm used in the standard likelihood model (Tarantola 2005). It measures the deviation from the perfectly correlated trace normalized by the variance, which reflects the level of confidence in the evaluated correlation coefficient. Thus, note, that highly similar synthetic and real seismic reflection data (i.e., a correlation coefficient close to 1) receive the a misfit score ($M$) close to zero.

In this work, we introduce an approach for multi-scale uncertainty assessment in geostatistical seismic inversion. It combines the local uncertainty, as assessed by conventional iterative global acoustic geostatistical seismic inversion, namely the Global Stochastic Inversion (GSI; Soares et al. 2007), and the large-scale geological uncertainty, represented by a spatial continuity model (i.e., a variogram model) and the prior probability distributions of petro-elastic reservoir properties. The proposed approach uses adaptive stochastic sampling, namely Particle Swarm Optimization (PSO), and Bayesian inference to quantify their posterior probability (Mohammed et al. 2010) (Figure 1). The method is illustrated with application examples to both synthetic and real datasets using fullstack seismic reflection data to infer acoustic impedance (Ip) models.

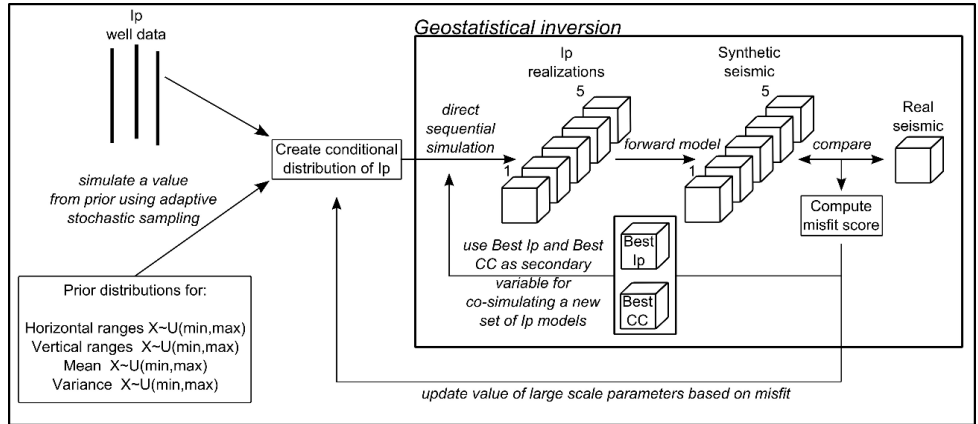

Figure 1 –Schematic representation of the proposed methodology to include multi-scale uncertainty assessment in geostatistical seismic inversion.

## METHODOLOGY

We address the problem of assessing the uncertainty related to inverted petro-elastic properties of the subsurface at hierarchically different, but complementary, levels: (i) local uncertainty at the grid cell scale related to the subscale heterogeneity; (ii) large-scale geological uncertainty related to spatial correlation and global distribution of the petro-elastic properties of interested. Local uncertainty is inferred by stochastic sequential simulation, used as the perturbation technique of the model parameter space in the geostatistical seismic inversion procedure. Large reservoir scale geological uncertainty is inferred through a Bayesian framework, using stochastic adaptive sampling based on the prior uncertainty related to the spatial correlation parameters and the prior range of the global distribution of the petro-elastic properties of interest. Figure 1 summarizes the proposed workflow for assessing multi-scale uncertainty during geostatistical acoustic seismic inversion.

### Geostatistical seismic inversion

The local uncertainty is assessed by the intrinsic nature of the iterative geostatistical seismic inversion algorithm used to invert seismic reflection data for elastic properties. At each iteration, during the inversion process, a set of *Ns* impedance models is created at once for the entire inversion grid using stochastic sequential simulation and co-simulation (Soares 2001). This allows assessing the variability between the simulated models (or realizations) for each node of the inversion grid. All the realizations are equivalent under the same a priori assumptions on the prior probability distributions of Ip and on its spatial continuity pattern. The variability on each grid node, among an ensemble of realizations, is achieved by using different simulation paths at each realization and the Monte Carlo sampling from the local pdf. The conditioning data is defined as the combination of the available experimental data and pre-simulated nodes within a pre-defined neighborhood, which depends on the spatial correlation model. The changes on the conditioning data for the same grid node between different realizations are due to the use of a distinct random paths for visiting and simulating all the nodes within the simulation grid for each simulation run. Since the grid nodes are visited and simulated in a different order from realization to realization the pre-simulated nodes in the conditioning data will also be different allowing for distinct simulated values among the set of realizations (Deutsch and Journel 1998).

Uncertainty assessed through conventional geostatistical seismic inversion methodologies is associated with each separate seismic trace across the entire inversion grid, therefore reflects more local aspects rather than global at larger geological scales. This bears some limitation regarding uncertainty assessment related to larger scale geological features, such as stratigraphic structure, geological continuity, facies proportions, etc. These features are related to the entire reservoir and cannot be accurately inferred from a single trace.

The uncertainty of the large-scale geological parameters of the geostatistical seismic inversion is inferred in a Bayesian way from the ensemble of matched (high likelihood) models generated by adaptive stochastic sampling (e.g., particle swarm optimization). The posterior probability of the high likelihood models is then approximated using a Neighborhood Algorithm Bayes (NAB, Sambridge 1999)). Marginal posterior probability distributions for each parameter can be derived from the approximated PPD. The parameter uncertainty is related to the underlying global geological continuity described by the variogram model and to the prior global probability distribution of Ip inferred from the existing Ip well-log data. Uncertainty in spatial continuity accounts for the variation in the horizontal and vertical ranges of the variogram model and the azimuth angle. Uncertainty in the prior probability distribution of Ip is represented by the varying shape of distribution parameterize with a Gaussian Mixture model (GMM). The GMM is defined by the means (μ) and standard deviations ($\sigma_{GMM}$) of $k$ Gaussian modes, where each mode corresponds to a different facies. The a priori choice of the GMM is done with respect to the existing well-log data and geological analogues. The GGM distribution used as conditioning data of the stochastic sequential simulation of Ip is generated by inverse cumulative transform using the sampled μ and $\sigma_{GMM}$.

At the end of each iteration the uncertain parameters are updated based on the likelihood, calculated based on the misfit-score (Equation 3) between the best inverted seismic data at the end of each iteration of the geostatistical seismic inversion loop and the real seismic reflection data.

**Inference of the large-scale geological parameters**

In the conventional approach, geostatistical seismic inversion methodologies, such as the GSI (Soares et al. 2007), assume that the spatial continuity pattern of Ip as modelled from the experimental variogram, computed from existing well data, is known and with no uncertainty for the entire inversion grid. This is a very broad assumption and is hardly true for real case studies with only few sparsely drilled wells available at early exploration stages. In those particular cases, the calculation of horizontal experimental variograms is not very accurate, with a high degree of interpretational uncertainty, due to the limited number of existing experimental data values during its calculation (Figure 2). The global a priori probability distribution of Ip is also usually inferred directly from the available well-log data, subject to preferential well placement, with some respect to available analogue and geological outcrop data. Stochastic sequential simulation algorithms used in these geostatistical seismic inversion methodologies are designed to accurately reproduce the target a priori probability conditioning distribution function (Deutsch and Journel, 1998). In the specific case of GSI, DSS is used as the model perturbation technique. This stochastic sequential simulation methodology was designed to increase the accuracy in the reproduction of complex multimodal target distributions (Soares 2001). Furthermore, the use of a global probability distribution estimated exclusively from available well-logs as a target for the stochastic sequential simulation algorithm remains a limitation of the iterative geostatistical inversion algorithm. The uncertainty related to the estimation of this distribution is mainly related to the bias in the wells location. Commonly, the wells are drilled in sand-prone areas (often associated with low impedance values) and, therefore, the data from them are biased towards the reservoir sweet spots. This creates a difficult in accurate determination of the target reservoir property distribution for less porous lithology. Uncertainty of the target global distribution still needs to be addressed through the statistical inference. The target distribution based on the well data can be parameterized and then its parameters are inferred in an inverse way using seismic reflection data.

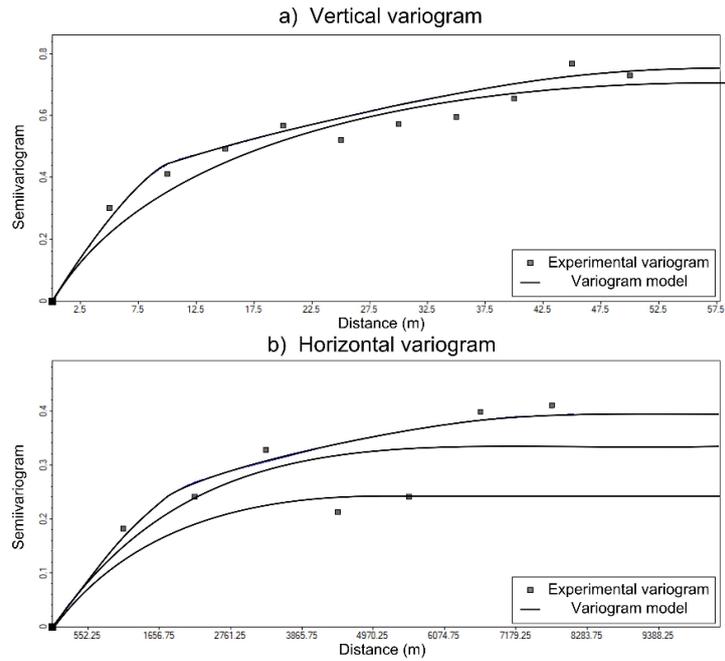

Figure 2 – Uncertainty on modelling experimental vertical and horizontal variograms. Several variogram models with different correlation ranges and structure can be fitted to the same experimental variograms depending on the geomodeller's preference and experience. (a) experimental vertical variogram and (b) experimental horizontal variogram, both computed from Ip-logs of all wells of the real case application example.

The proposed workflow links the Bayesian inference with the geostatistical inversion workflow in order to optimize the referred GSI parameters allowing assessing the large-scale geological uncertainty along with the local uncertainty due to small scale heterogeneity inferred from the GSI. Note that the same concepts apply to the different flavors of global iterative geostatistical seismic inversion methodologies that are able to invert seismic reflection data for acoustic impedance (Soares et al. 2007), for acoustic and elastic impedance (Azevedo et al 2015) or simultaneously for P- and S-wave velocities and density (Azevedo et al. 2017). The application examples shown in this work comprise the use of fullstack seismic reflection data but its extension for the elastic domain is straightforward.

## APPLICATION EXAMPLES

We have applied the proposed methodology to both synthetic and real case studies. The application of this methodology to the synthetic dataset demonstrates a proof of concept, since we have access to the ground truth solution, while the real case application shows the robustness of the proposed approach in the presence of few well data and realistic noise level. The real case study results are validated towards real wireline log data from wells not used to constrain directly the inversion procedure (i.e., blind wells), compared against those retrieved from a typical geostatistical seismic inversion.

**Synthetic application**

The Stanford VI dataset (Castro et al. 2005) was used as the synthetic dataset. The dataset includes a set of petro-elastic models, created by geostatistical algorithms, accompanied by the corresponding synthetic 3D seismic reflection data. We have selected one of the three original reservoir sections corresponding to the meandering channels sedimentary environment. The meandering geological pattern of sand channels can be easily identified in horizontal sections extracted from both the true acoustic impedance model and the true seismic reflection data (Figure 3).

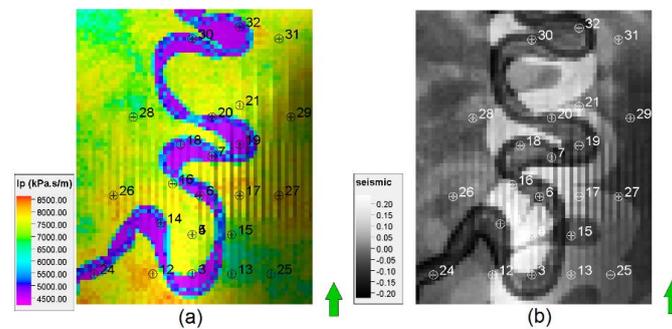

Figure 3 – Horizontal sections extracted from (a) the true acoustic impedance model and (b) the true seismic reflection data and available well dataset from Stanford VI.

The proposed approach was applied to a noise-free seismic volume computed from the true acoustic impedance model, by convolving the reflection coefficients, derived from the true Ip model, with a wavelet. The same wavelet used to compute the reference fullstack volume was also used as part of the inversion procedure (i.e., no uncertainty related to the wavelet was taken into account). The inversion grid has 60 by 70 by 20 cells in the *i*-, *j*- and *k*-directions respectively, and is conditioned by 23 wells with Ip-log information (Figure 3).

The global Ip distribution estimated from the available well-log data does not quite capture the relative proportion of each population and the minimum and maximum Ip values as inferred from the true Ip (Figure 4, Table 1). In real datasets, it is expected these differences to be more prominent due the preferential well placement along sand-prone facies.

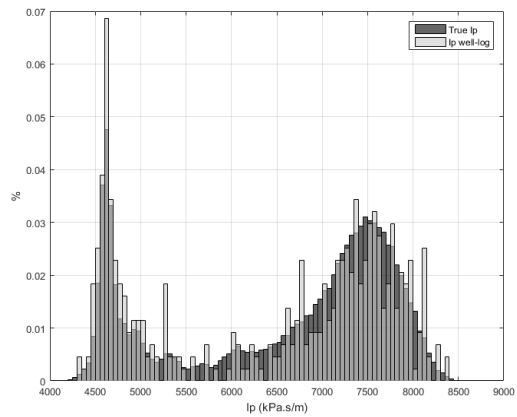

Figure 4 – Comparison between distributions estimated from the true Ip and the Ip-log data from the twenty-three existing wells.

Table 1 – Main statistics computed from the true Ip and the existing Ip-log data.

|  | Mean (kPa.s/m) | Variance ((kPa.s/m)2) | Min. (kPa.s/m) | Max (kPa.s/m) |
|---|---|---|---|---|
| **True Ip** | 6586.29 | 1483469.28 | 4218,36 | 8632,04 |
| **Ip well-log** | 6424.13 | 1723261.69 | 4327,90 | 8391,23 |

The large-scale geological uncertainty was inferred by varying simultaneously the horizontal and vertical ranges of the variogram model used as part of the stochastic sequential simulation and co-simulation, and the global target probability distribution used as the control parameter in the sequential simulation. The uncertainty of the global target probability distribution was represented by the mixture of two Gaussian models, defined by a plausible range for their means and standard deviations (Table 2).

Table 2 – Summary of the a priori parameterization of the large-scale geological variables optimized during the iterative procedure.

| Parameter | Type of distribution | Distribution |
|---|---|---|
| Horizontal variogram range (m) | Uniform | 700-800 |
| Vertical variogram range (m) | Uniform | 10-200 |
| Mean Gaussian 1 (kPa.s/m) | Uniform | 3500-4500 |
| Mean Gaussian 2 (kPa.s/m) | Uniform | 4500-7500 |
| Variance facies 1 (Sigma Gaussian 1) ((kPa.s/m)$^2$) | Uniform | 200000-300000 |
| Variance facies 2 (Sigma Gaussian 2) ((kPa.s/m)$^2$) | Uniform | 70000-90000 |
| Proportions facies 1 (proportion 1) (%) | Uniform | 30-70 |

At each iteration, five Ip models were generated using the inferred large-scale geological parameter values. A long run of 500 PSO iterations (Figure 5) was performed in order to ensure the convergence of the adaptive sampling in each parameter, as it homes into the intervals of low misfit, with the misfit as defined in Equation 2. After the first 200 iterations the misfit decreases and the procedure can be considered converged. The remaining iterations generated consistently Ip models with low misfit (Figure 5). Several high misfit models towards later iterations occur due to the stochastic nature of the proposed procedure, i.e. the exploration of the model parameter space with the introduction of a new generation of particles to avoid an overfitting of the procedure and the intrinsic stochasticity of geostatistical simulations.

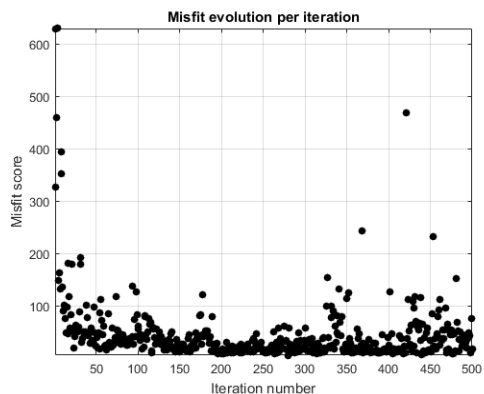

Figure 5 – Misfit evolution versus iteration number for the multi-scale uncertainty assessment in seismic inverse methodologies of the synthetic example.

Figure 6 shows the evolution of inverted Ip model at two moments along the iterative procedure. At the second iteration, the Ip model is still very discontinuous and the spatial distribution of the inverted Ip-models is far from the true one. On the other hand, the mean model of the five realizations corresponding to the lowest misfit score (Figure 6c) shows an Ip model closer to the true Ip, and able to retrieve the non-stationary shape of the main turbidite channel. The same evolution pattern can be interpreted from the inverted synthetic seismic data (Figure 7) generated from the Ip mean models shown in Figure 6.

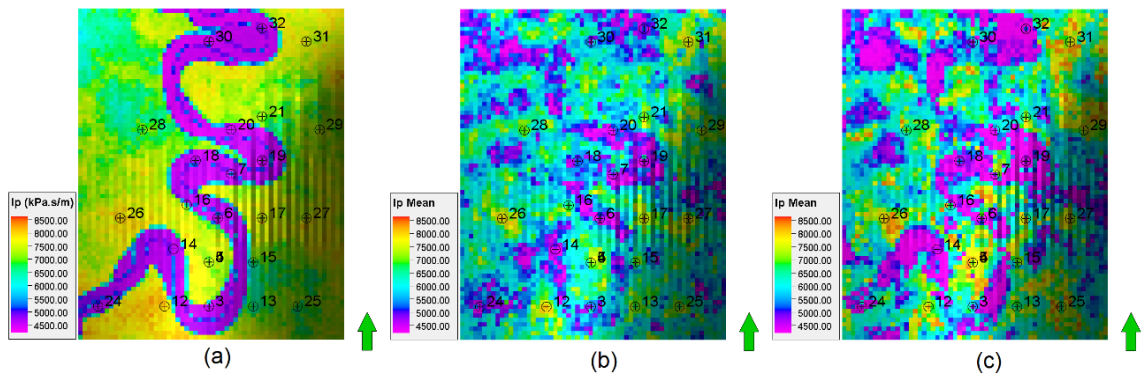

Figure 6 – Horizontal slices extracted from: (a) the true acoustic impedance; (b) the mean model of the acoustic impedance realizations generated at the second iteration; and (c) the mean Ip model computed from the iteration with lowest misfit.

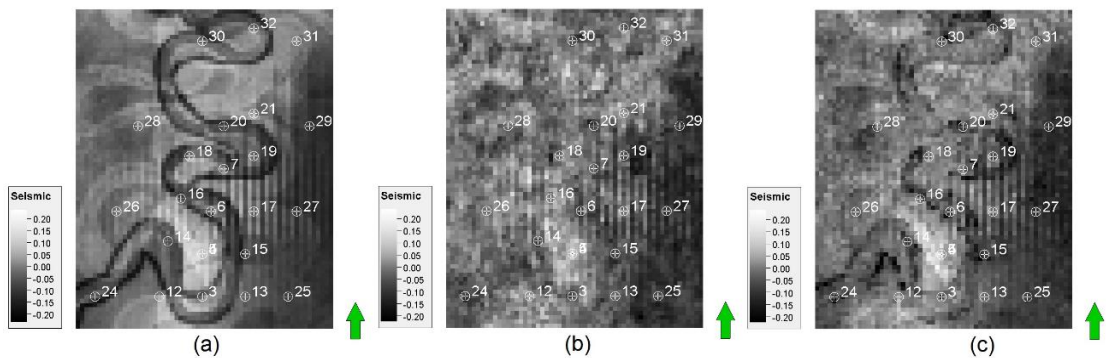

Figure 7 – Horizontal slices extracted from: (a) the real seismic data; (b) the synthetic seismic generated from the mean Ip model computed at the second iteration; and (c) the synthetic seismic volume generated from the mean Ip model computed from the iteration with lowest misfit.

Besides the evolution of the Ip models generated during the iterative procedure it is of interest to assess how the parameters related to the large-scale geological features evolved. Figure 8 shows the evolution of these seven parameters. At the beginning of the iterative procedure, all parameters explore the a priori range of uncertainty as characterized by uniform distributions ranging between the user-defined limits (Table 2). As the iteration procedure evolves, the exploration of the parameter space homes in to the values which indicate the posterior uncertainty related to the specific parameter.

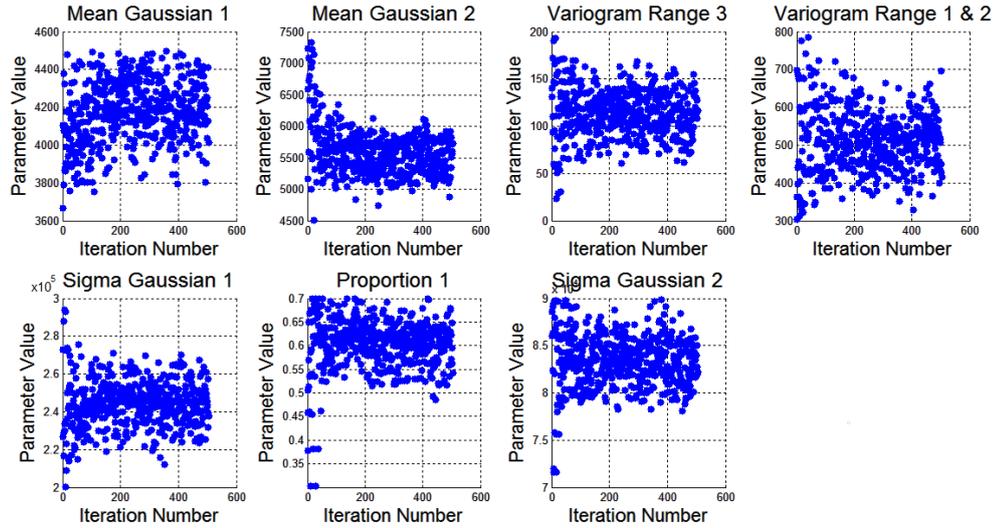

Figure 8 – Parameter evolution with iteration number. Globally it shows a reduction in the uncertainty range for all parameters. The sampling of the parameter space narrows while the number of iterations increases. Horizontal variogram ranges (variogram range 1 and 2) are expressed in meters while the vertical variogram range (variogram range 3) is defined in milliseconds.

The evolution of the parameter values and the homing towards a smaller posterior interval through the iterations leads to the decrease in the misfit score (Figure 9). The range of simulated parameter values, corresponding to the smaller misfit score, represents the parameter uncertainty and is related to the non-unique solution of seismic inversion problems.

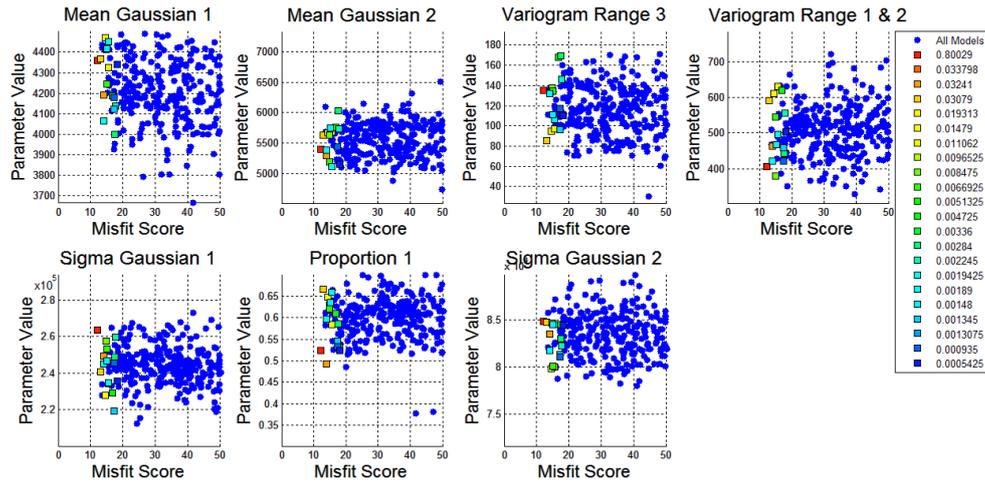

Figure 9 – Parameter evolution with misfit score and colored by its posterior probability of occurrence after Bayesian inference. Horizontal variogram ranges (variogram range 1 and 2) are expressed in meters while the vertical variogram range (variogram range 3) is defined in milliseconds.

The marginal PPD for each parameter is derived from the approximated PPD based on the set of 500 models generated during the iterative procedure. Figure 9 depicts the parameter values and the corresponding likelihood for the models resampled to infer the PPD. The colour code depicts the range of the approximated posterior probability. These models can be used to assess and quantify the uncertainty related to each large-scale geological parameter, by computing the posterior credible intervals. Also, a maximum a posterior (MAP) value for each parameter is shown that relates to the model with the highest posterior probability.

**Real case application**

The proposed technique was applied to a real onshore field, where the available dataset comprises a fullstack volume, five wells with Ip-logs (Figure 10) and a wavelet extracted from the seismic volume and calibrated at the well locations (Azevedo et al. 2014). No uncertainty related to the wavelet extraction procedure was incorporated in this application example. The entire volume is composed by 350x198x49 cells in the $i$-, $j$- and $k$-direction respectively. In order to assess the robustness of the proposed method only one of the five wells (W1) was kept as conditioning data for the generation of Ip models with stochastic sequential simulation during the iterative inversion procedure. The remaining four wells were used exclusively as blind wells to compare the inverted Ip values against the true Ip-logs and assess the performance of the proposed method.

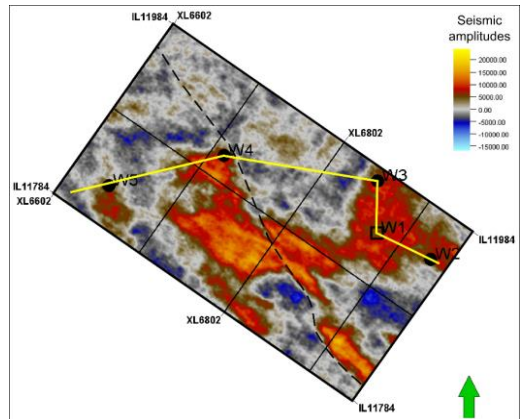

Figure 10 – Inversion grid and relative location of the available wells used for the multi-scale uncertainty assessment in geostatistical seismic inversion. Well W1 was the only well used as the conditioning data for the geostatistical seismic inversion. Yellow line shows the location of the well cross-section presented in the Figure 13. Black dashed line represents the interpretation of a fault that compartmentalizes the reservoir.

The Ip distribution as inferred exclusively from well W1 is not able to capture the variability of the true Ip-logs using the existing five wells (Figure 11). In a conventional geostatistical seismic inversion approach, all the Ip models generated during the inversion procedure would reproduce the distribution as expressed by W1, showing a bias when compared against the true Ip distributions and resulting in geological inconsistent subsurface models.

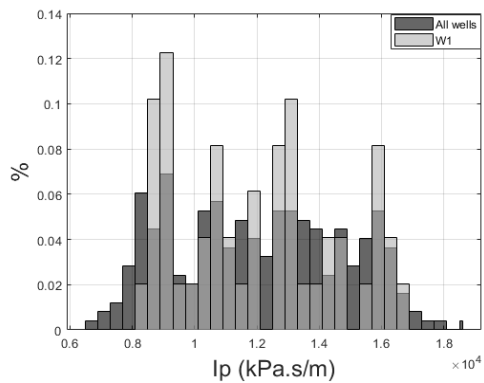

Figure 11 – Comparison between the histograms inferred from the entire set of Ip-logs and from well W1 exclusively. Nor the minimum and maximum values of Ip are comprehended in the limits inferred from W1.

In the real example we explored the model parameter space for ten large-scale geological parameters related to the spatial continuity pattern of Ip and its distribution values (Table 3). The a priori ranges were defined by the geological knowledge of the study area and true Ip-logs available for the five wells. In this specific case, three different facies (i.e., Gaussian mixtures) were selected to approximate the observed Ip distribution and the spatial continuity pattern of Ip was inferred by optimizing the horizontal and vertical ranges of a single variogram model (Table 3).

Table 3 – Summary of the a priori parameterization of the large-scale geological variables optimized during the iterative procedure for the real case study.

| Parameter | Type of distribution | Range |
|---|---|---|
| Horizontal variogram range (a1, m) | Uniform | 250-2500 |
| Vertical variogram ranges (a3, m) | Uniform | 8-40 |
| Mean facies 1 (mfac1, kPa.s/m) | Uniform | 8000 - 12000 |
| Mean facies 2 (mfac2, kPa.s/m) | Uniform | 9000 - 13000 |
| Mean facies 3 (mfac3, kPa.s/m) | Uniform | 12000 - 17000 |
| Std. dev. facies 1 (sigfac1, kPa.s/m) | Uniform | 1400 - 1600 |
| Std. dev. facies 2 (sigfac2, kPa.s/m) | Uniform | 1700 - 1900 |
| Std. dev. facies 3 (sigfac3, kPa.s/m) | Uniform | 1400 – 1600 |
| Proportions facies 2 (pfac2, %) | Uniform | 30 - 40 |
| Proportions facies 3 (pfac3, %) | Uniform | 30 - 40 |

For the sake of comparison, a conventional geostatistical acoustic inversion (GSI; Soares et al. 2007) was ran using W1 as constraining data. The inversion procedure was parameterized with six iterations where ensembles of thirty-two realizations were generated at each iteration using direct sequential simulation and co-simulation (Soares 2001). The variogram model imposed for the stochastic sequential simulation and co-simulation was eye-fitted from experimental variograms computed from all the five wells. The best-fit impedance model generated during the iterative procedure (Figure 12a) produces synthetic seismic reflection dataset with a global correlation coefficient of 97.4% when compared with the real fullstack volume. In fact, all the thirty-two models (Figure 12b and Figure12c) generated during the last iteration result in synthetic seismic dataset with a correlation coefficient above 97% when compared against the real fullstack volume. The variability between realization is small (Figure 12). Despite the good global correlation coefficient between the observed and the real seismic reflection data, the best-fit Ip model lacks the reproduction of the measured data at the wells locations not used as conditioning data. This effect is more pronounced at shallower depths (between the 1805ms and 1950ms) around wells W4 and W5, which are far from well W1.

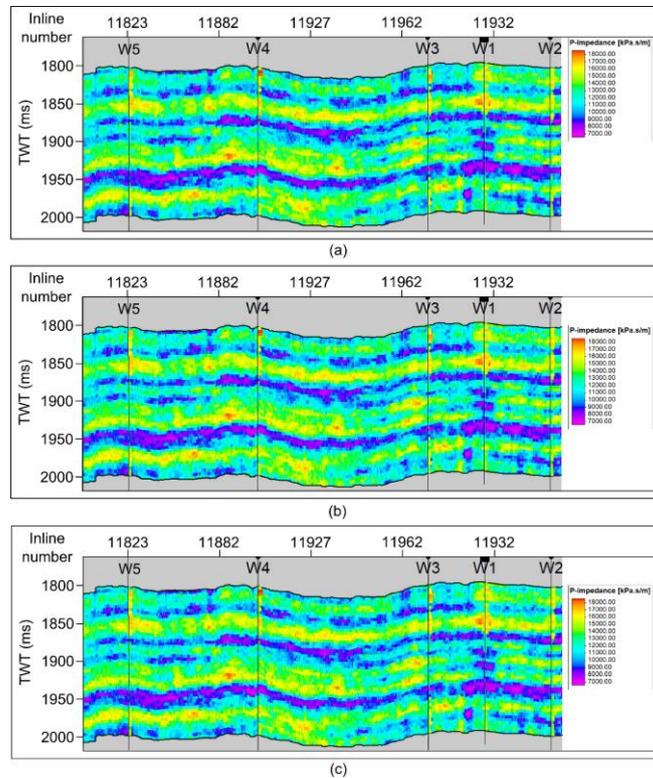

Figure 12 – Vertical well-section extracted from: (a) the best-fit model of Ip inverted using geostatistical acoustic inversion conditioned to well W1; (b) realization number thirty-one from last iteration; and (c) realization number one from last iteration.

Figure 13 shows the comparison between the real Ip-log and the Ip traces inverted during the last iteration of the geostatistical inversion at the blind well locations. The ensemble of 32 GSI realizations from the last iteration underestimates the uncertainty. While globally the match between inverted and true logs is acceptable, it is also possible to interpret that the inverted traces lack reproducing the measured Ip-log, as the increasing of Ip from 3200 to 3300 m. Besides these discrepancies, the synthetic seismic traces generated from these elastic traces do have a high correlation coefficient when compared against the collocated real seismic traces. It is also clear, that all the inverted Ip traces generated at this iteration are similar and are not able to encompass the real Ip-log. We may interpret it as a lack of the exploration of the model parameter space, where the inversion procedure is trapped at a local minimum far from the true solution.

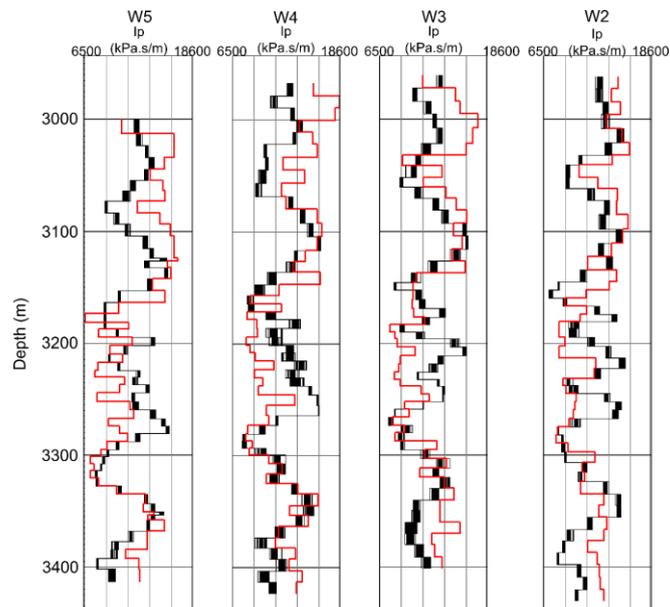

Figure 13 – Well section compares the Ip-log data for all the wells (red line) against the inverted Ip traces from the thirty-two realizations of Ip generated during the last iteration and shows underestimation of uncertainty.

We ran a long 450 iteration of the proposed method. On each iteration, the values of the large-scale geological parameters are updated. Per iteration, five realizations of Ip are generated with stochastic sequential co-simulation (Figure 14). As for the synthetic example, the use of such a large number of iterations ensures the convergence of all the parameters selected to be optimized. The misfit score decreases gradually along the 450 iterations run. There is a considerable decrease on the misfit value for the first fifty iterations, after which it decreases slowly. At iteration 325 a new set of particles was introduced in the procedure to avoid overfitting and ensure a wider exploration of the model parameter space.

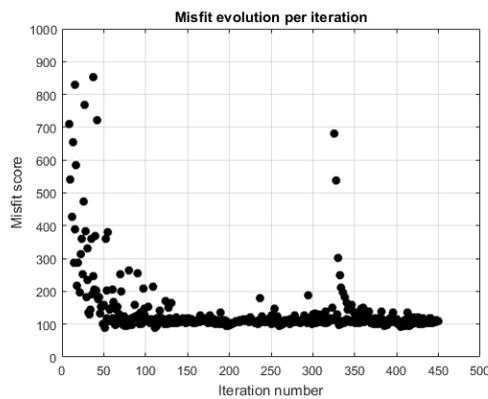

Figure 14 - Misfit evolution versus iteration number for the multi-scale uncertainty assessment in seismic inverse methodologies for the real case application.

During the long optimization run the value of each parameter converged gradually towards a narrower interval of values while ensuring the minimization of the misfit value (Figure 15). It is worthwhile noting that not all the parameters converged with the same convergence ratio, allowing to infer which parameters are more uncertain and of relevance to the problem at hands. When comparing the parameters related to facies distribution, those related to the definition of facies 2 show a broader range of possible values at the end of the optimization. This is the expected behavior when looking to the Ip distribution as inferred from W1 (Figure 11). Facies 2 corresponds to the intermediate values of Ip, while there is only one mode in the distribution as revealed by well W1, it is also possible to interpret two different modes within the same range of values if all the five wells are considered (Figure 11). At the end of the iterative procedure, the posterior distribution of the parameters related to facies 2 represent the ambiguity of expressing two different populations with a single mixture.

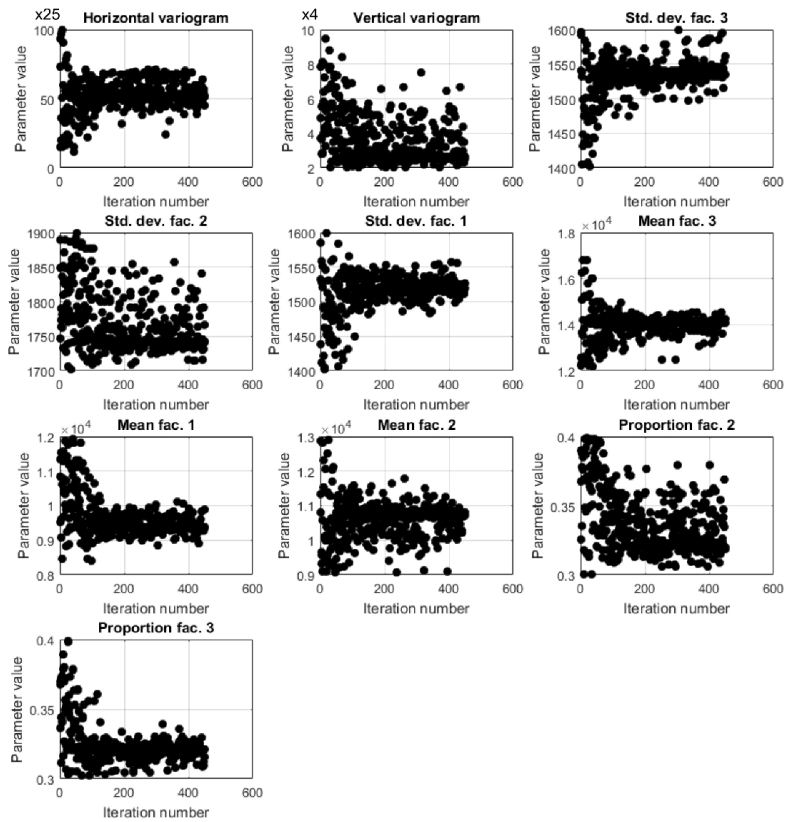

Figure 16 – Parameter evolution with misfit score, colored by its probability of occurrence after Bayesian inference. Horizontal variogram ranges (variogram range 1 and 2) are expressed in meters while the vertical variogram range (variogram range 3) is defined in milliseconds.

The set of 450 iterations was then used to infer its probability of occurrence using the NAB algorithm (Figure 16). In this real case application one single model (i.e., the best-fit inverse model) takes approximately 57% of probability of occurrence due to the much lower value of misfit for this specific model when compared with the remaining. The models resampled during the NAB were used to estimate the P10, P50 and P90 Ip distributions (Figure 17). In what concerns the large-scale features, these models do agree with the best-fit inverse model retrieved from the conventional GSI (Figure 12). The main differences are related to the small-scale variability and the spatial continuity of the main events: the models retrieved by the proposed methodology are smoother and show higher horizontal continuity. This may allow interpreting that the horizontal variograms estimated from the five wells, and imposed as part of the GSI, contradict the information provided by the seismic reflection data and captured by the proposed methodology by using an a priori range rather than a single value.

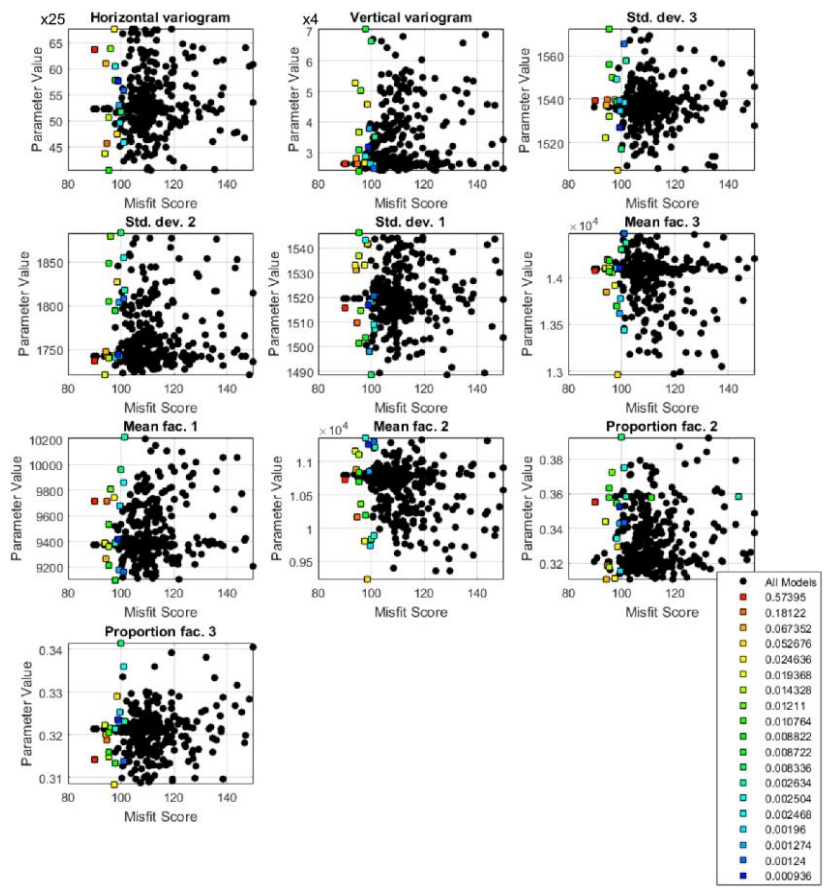

Figure 16 – Parameter evolution with misfit score, colored by its probability of occurrence after Bayesian inference. Horizontal variogram ranges (variogram range 1 and 2) are expressed in meters while the vertical variogram range (variogram range 3) is defined in milliseconds.

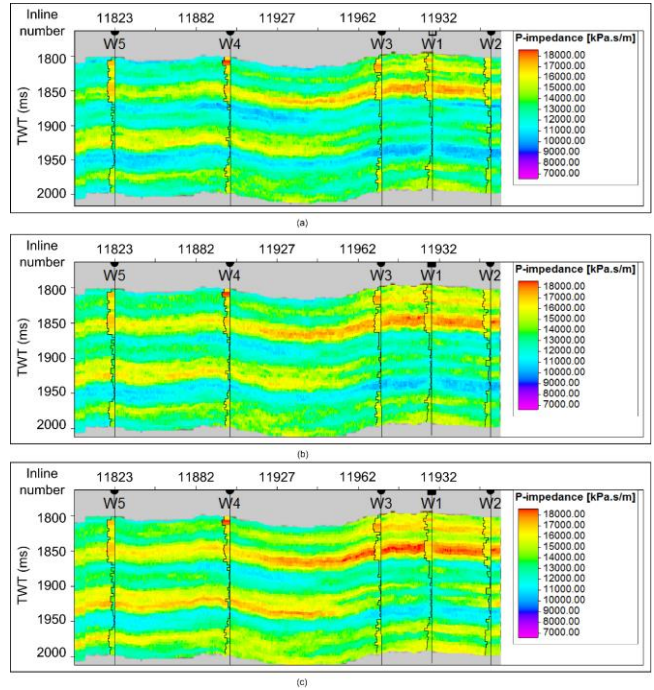

Figure 17 – Vertical well-section extracted from: (a) P10; (b) P50; and (c) P90 Ip models.

The blind well locations were used to assess: how good the proposed methodology is able to explore the model parameter space and how the prior distribution for each parameter allows encompassing the measured Ip log (Figure 18); and how the estimated uncertainty envelope encompasses the true Ip as revealed by the existing well-log data (Figure 19).

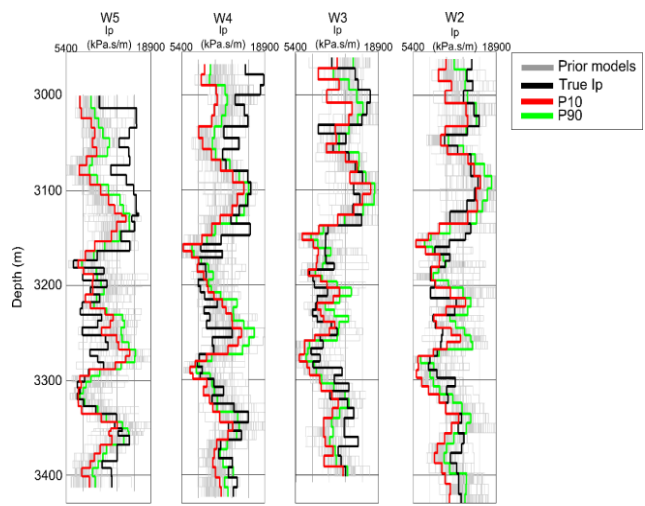

Figure 18 – Vertical well section for blind wells showing the prior distribution of Ip as represented by Ip models generated at the first iteration of the iterative inversion

procedure without being conditioned to the seismic data and the P10 and P90 models estimated from the ensemble of models resampled by the NAB.

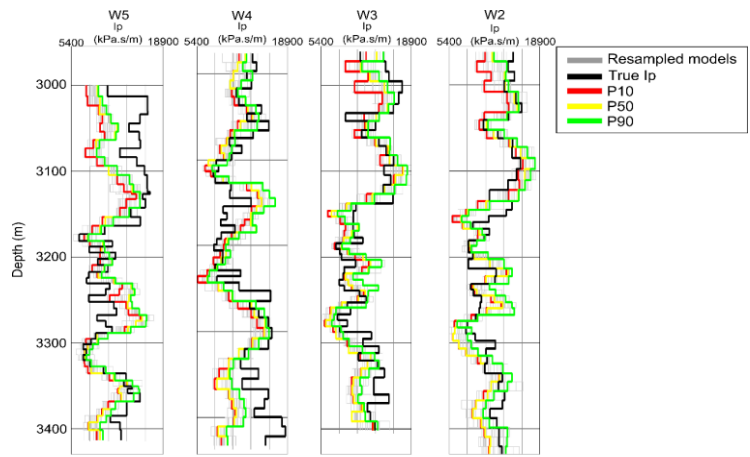

Figure 19 – Comparison between the real Ip log at the blind well locations with all the posterior models resampled by the NAB and the P10, P50 and P90 models computed from this ensemble.

The set of Ip realizations generated at the initial steps of the iterative procedure (i.e., only conditioned by the W1) using large-scale geological parameters sampled by the adaptive stochastic sampling algorithm are able to capture the measured Ip log, considered as the true subsurface Ip, for wells W3 and W2. At shallower depth, the selected prior distributions are not able to encompass the true Ip for well W4 and W5. These discrepancies are related to the different geological context (different sedimentary environment as represented by two distinct fault blocks) where these two wells rely when compared with wells W1, W2 and W3 (Figure 10) (Azevedo et al. 2014).

Bayesian credible P10-P90 interval has been derived based on the resampled models to assess how well the uncertainty associated with the large-scale geological parameters is captured. The posterior quantiles are derived from the posterior ensemble of the Ip cubes with respect to approximated PPD. This becomes possible because each Ip volume is associated with the corresponded PPD value. Therefore, a local conditional PPD can be obtained for any grid cell. The Ip P10, P50 and P90 were computed for each trace location from the corresponding PPD (Figure 20).

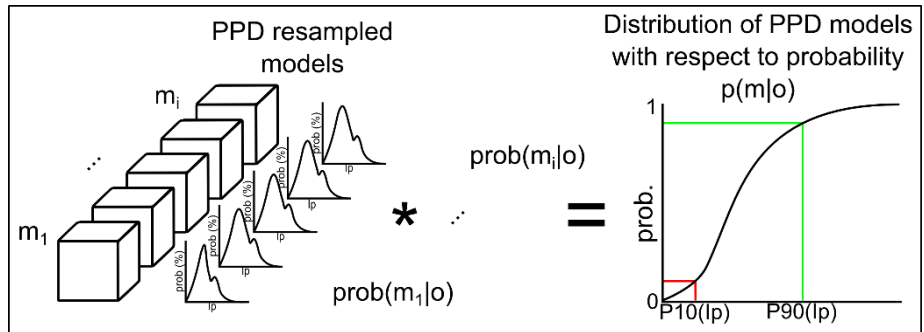

Figure 20 – Schematic representation of how the Bayesian credible P10-P90 interval is inferred from the PPD resampled models.

Figure 18 compares the estimated P10 and P90 with the true Ip and the range of the prior models. The Ip credible interval does approximate better when compared against the span resulting from the conventional geostatistical acoustic inversion, though it is not able to completely incorporate the true measurements (Figure 13). The use of the proposed method increased by 21% the number of samples of the true Ip-logs within the P10- P90 envelope. In the cases when the true Ip deviates from the P10-P90 envelope it is also outside the prior range. This suggests underestimation of the prior uncertainty and, thus, more uncertain parameters may be possibly considered.

The proposed methodology to explore in a more comprehensive way the model parameter space when compared with the conventional approach. This aspect can be critical for the success of reservoir modeling and characterization projects and better assess the risk associated with a given decision.

Finally, to assess the ability of the proposed methodology in reproducing the Ip distribution as inferred simultaneously from all the five wells we compared the distribution of Ip-logs with the one retrieved from the P10, P50 and P90 models (Figure 21). Contrary to the conventional case, where the Ip distribution is always conditioned to the one inferred from the values of Ip available for well W1, it is possible to interpret that the uncertainty enveloped represented by the interval between the P10-P90 models are able to get a good estimate of the distribution as inferred from all the wells, including those used exclusively as blind tests.

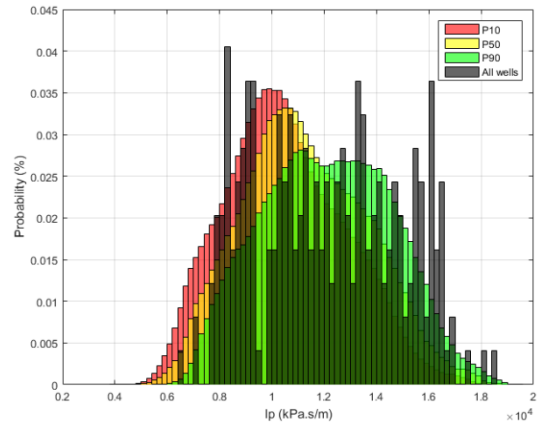

Figure 21 - Comparison between the histograms computed from the real Ip logs for all wells available (including the blind well tests) and the P10, P50 and P90 models.

## DISUCUSSION

The results presented for both the real and synthetic case applications are consistent and justify the relevance of the proposed approach in geostatistical seismic reservoir characterization. They show how uncertainty in the geological description of the reservoir properties can be assessed and rigorously integrated in geostatistical seismic inversion.

The application examples show that the prior distribution ranges functions for each parameter are of extreme importance in the assessment and quantification of the uncertainty related to these parameters. The synthetic example has demonstrated a more reliable uncertainty envelope due the larger amount of the conditioning well data and more informed prior distributions, which appeared to be a better reflection of the idealized (Gaussian-based) synthetic seismic. The real case application is more challenging because the seismic reflection is not noise-free and there are uncertainties related to the wavelet estimation that are not taken into account under this study. Moreover, Figures 18 and 19 show that for wells W4 and W5 we are not able to properly infer the Ip log measured at the shallower depths of the wells. These discrepancies are related to the lack of a reliable definition and flexibility of the prior distributions for the Gaussian mixtures used to approximate the Ip distributions. In fact, Figure 18 show that the prior ensemble of models does not capture the natural variability as interpreted from the well log, which suggests more uncertainties to be considered a priori. The generation of these models, and consequently how is their spread, is conditioned to the number and location of well data used and the type of variogram model imposed.

Nevertheless, the GSI application to the real case reveals the lack in reproducing the measured Ip-logs while the span of the inverted elastic traces at the end of the iterative procedure is small (Figure 13). This is due to the reproduction of a fixed target probability distribution, in this case as inferred exclusively from a single well W1, and a stationary spatial continuity pattern, assumed known in the stochastic sequential simulation technique.

## CONCLUSIONS

We introduce herein a way to quantify uncertainties on both large geological scale and small grid resolutions scale by coupling geostatistical seismic inversion coupled with adaptive stochastic sampling and Bayesian inference of uncertain parameters associated with the geology. The latter uncertainty are traditionally overlooked in geostatistical seismic inversion techniques and these parameters, such as a variogram model and global distributions, are assumed known. In geostatistical seismic inversion, the uncertainty is normally related to the variability of the spatial distribution of the elastic property of interest and represented by a statistical measurement computed from an ensemble of models generated at a given iteration (e.g., the variance within the ensemble generated at a given iteration). By coupling stochastic adaptive sampling and Bayesian inference we were able to include uncertainty in large-scale geological parameters that express the spatial continuity pattern of the property of interest and the uncertainty related to the bias that exists due to the lack the limited set of existing experimental data.

The proposed methodology applied to a synthetic and real case demonstrated a more adequate uncertainty quantification with the inferred credible P10-P90 envelope, which agrees better with the true seismic in the blind wells that the one from the traditional GSI. This opens provides new insights towards the integration of different layers of uncertainty in geostatistical seismic inversion techniques. While the examples shown here only consider the acoustic case, the extension of the proposed technique to the elastic domain is straightforward.

## ACKNOWLEDGEMENTS

The authors acknowledge Epistemy for the donation of academic licenses of Raven® and Schlumberger for the donation of Petrel®. LA gratefully acknowledge the support of the CERENA (strategic project FCT-UID/ECI/04028/2013).